\documentclass[conference]{IEEEtran}
\usepackage{amsmath,graphicx}
\usepackage{psfrag}
\usepackage{pgfplots}
\usepackage{balance}

\pgfplotsset{compat=newest}
\pgfplotsset{plot coordinates/math parser=false}
\newlength\figureheight
\newlength\figurewidth

\usetikzlibrary{plotmarks}

\newcommand{\argmax}{\mathop{\mathrm{argmax}}}

\title{Adaptive Frequency Prior for Frequency Selective Reconstruction of Images from Non-Regular Subsampling}

\author{\IEEEauthorblockN{J{\"u}rgen Seiler and Andr{\'e} Kaup}
\IEEEauthorblockA{Multimedia Communications and Signal Processing\\
		Friedrich-Alexander-Universit{\"a}t Erlangen-N{\"u}rnberg (FAU), Cauerstr. 7, 91058 Erlangen, Germany\\
		\fontsize{10}{10}\selectfont\ttfamily\upshape
		\,juergen.seiler@FAU.de,  
		\,andre.kaup@FAU.de}
}

\begin{document}


\maketitle

\begin{abstract}
Image signals typically are defined on a rectangular two-dimensional grid. However, there exist scenarios where this is not fulfilled and where the image information only is available for a non-regular subset of pixel position. For processing, transmitting or displaying such an image signal, a resampling to a regular grid is required. Recently, Frequency Selective Reconstruction (FSR) has been proposed as a very effective sparsity-based algorithm for solving this under-determined problem. For this, FSR iteratively generates a model of the signal in the Fourier-domain. In this context, a fixed frequency prior inspired by the optical transfer function is used for favoring low-frequency content. However, this fixed prior is often too strict and may lead to a reduced reconstruction quality. To resolve this weakness, this paper proposes an adaptive frequency prior which takes the local density of the available samples into account. The proposed adaptive prior allows for a very high reconstruction quality, yielding gains of up to 0.6 dB PSNR over the fixed prior, independently of the density of the available samples. Compared to other state-of-the-art algorithms, visually noticeable gains of several dB are possible.
\end{abstract}


\section{Introduction}
\label{sec:introduction}

Looking at image signals, it can be discovered that they are typically defined on a regular two-dimensional grid. This arrangement often directly results from the acquisition process of a digital camera. However, the positioning of the pixels on a regular grid also is very important for displaying, storing, and especially for processing the images. Nevertheless, there also exist scenarios where the amplitude information of an image is not available on a regular rectangular grid, but  only at non-regular positions. These positions can be regarded as a subset of positions with respect to a finer regular grid. 

The scenario where the pixels of an image are only available for a non-regular subset of positions might result from different reasons. For example, it may be caused directly by the acquisition process if techniques like the Optical Cluster Eye \cite{Meyer2011} or the Micro-Optical Artificial Compound Eyes \cite{Duparre2006} are used. Furthermore, a non-regular sampling can also be used intentionally for reducing the visible influence of aliasing \cite{Hennenfent2007, Maeda2009} or for enabling super-resolution techniques \cite{Schoeberl2011}.

Independent of the actual reason for the pixels of an image being only available at non-regular positions, for further processing or displaying the signals, a reconstruction on the full regular grid is required. This is necessary, since almost all signal processing algorithms rely on the input data being regularly spaced. As the available non-regularly spaced samples can be seen as a non-regular subsampling of the desired signal on the finer regular grid, the reconstruction task can be regarded as a resampling of the signal to a regular grid.

In literature, there exist many different algorithms for solving this under-determined task. This can be simple techniques, like Linear Interpolation, Four-Nearest-Neighbors Interpolation \cite{Ramponi2001}, or Natural-Neighbor Interpolation \cite{Sibson1981}. Other algorithms try to reconstruct a band-limited solution, given the non-regularly spaced samples \cite{Strohmer1997, Feichtinger1991, Marvasti1991}. Since the reconstruction is an under-determined problem, variation regularization based algorithms \cite{Dahl2010, Afonso2011} can also be used. Another class of algorithms exploits statistical properties of the underlying image signals \cite{Takeda2007,Zhai2012}. Aside from this, algorithms exploiting the fact that image signals can be sparsely represented in transform domains can be used \cite{Hawe2013, Elad2005}.

In \cite{Seiler2015}, we proposed the Frequency Selective Reconstruction (FSR), which is an effective sparsity-based algorithm for reconstructing image signals on a regular grid, given the samples at non-regularly spaced positions. As discussed in detail in \cite{Seiler2015}, FSR can also be seen within the Compressed Sensing framework and FSR is related to block-wise greedy Compressed Sensing reconstruction algorithms which make use of prior knowledge about the signal. One of the key features of the FSR is to account for the different likelihood of frequencies in image signals during the reconstruction process. For this, FSR incorporates properties of the optical transfer function of imaging systems.

However, we discovered that building up the frequency prior on the optical transfer function is a too strict constraint in some cases, leading to oversmoothing. Hence, in this paper we introduce a novel way to adaptively account for the prior knowledge about the frequency distribution. This prior adapts to the local density of available samples, leading to an improved reconstruction quality.

The paper is structured as follows. In the next section, the FSR algorithm is briefly revisited for providing a short overview of the algorithm and for showing the necessity of the adaptive frequency prior. Afterwards, the novel adaptive frequency prior is discussed in detail, before its effectiveness is shown in Section \ref{sec:results}. The paper ends with a short conclusion in Section \ref{sec:conclusion}.

\section{Brief Overview of Frequency Selective Reconstruction}
\label{sec:fsr_overview}

\begin{figure}
	\centering
	\psfrag{m}[l][l][1]{$m$}
	\psfrag{n}[l][l][1]{$n$}
	\psfrag{A}[l][l][1]{$\mathcal{A}$}
	\psfrag{B}[l][l][1]{$\mathcal{B}$}
	\psfrag{R}[l][l][1]{$\mathcal{R}$}
	\psfrag{L}[l][l][1]{$\mathcal{L}=\mathcal{A}\cup\mathcal{B}\cup\mathcal{R}$}
	\includegraphics[width=0.25\textwidth]{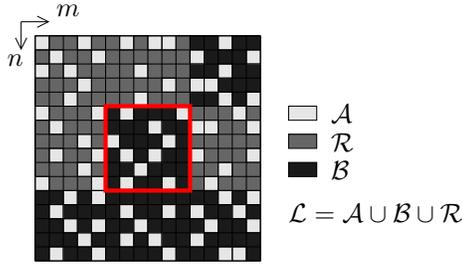}
	\caption{Reconstruction area $\mathcal{L}$ consisting of available samples in area $\mathcal{A}$, already reconstructed samples in area $\mathcal{R}$, and unknown samples in area $\mathcal{B}$. The currently processed block is marked in red in the center.}
	\label{fig:extrapolation_area}
\end{figure}

The Frequency Selective Reconstruction (FSR) as it is proposed in \cite{Seiler2015} is an effective algorithm for reconstructing image signals on a regular grid. For this, FSR performs a block-wise processing by dividing the image into equally-sized blocks. For reconstructing a block, it is always considered together with a spatial neighborhood, belonging to adjacent blocks. The pixels of a block and the neighborhood together form the extrapolation area $\mathcal{L}$. All samples in $\mathcal{L}$ can be divided into three groups as shown in Fig.\ \ref{fig:extrapolation_area}: known samples subsumed in area $\mathcal{A}$, unknown samples in area $\mathcal{B}$, and already reconstructed samples in area $\mathcal{R}$. Altogether, area $\mathcal{L}$ is of size $M\times N$ pixels.

The objective of FSR is to generate the parametric model 
\begin{equation}
 g\left[m,n\right] = \sum_{\left(k,l\right)\in\mathcal{K}}\hat{c}_{\left(k,l \right)} \varphi_{\left(k,l \right)}\left[m,n\right]
\end{equation}
as weighted superposition of Fourier basis functions $\varphi_{\left(k,l \right)}\left[m,n\right]$ for the samples in $\mathcal{L}$. There, $\hat{c}_{\left(k,l \right)}$ depict the estimated expansion coefficients and all used basis functions are subsumed in set $\mathcal{K}$.

The model generation works iteratively. In every iteration, one basis function is selected and its corresponding weight is estimated. Initially, the model $g^{\left(0\right)}\left[m,n\right]$ is set to zero and the residual $r^{\left(0\right)}\left[m,n\right]$ between the available samples and the model is equal to the available samples. For determining which basis function to add and for estimating its weight, in every iteration $\nu$ a weighted projection is carried out for calculating the projection coefficients
\begin{equation}
p_{\left(k,l \right)}^{\left(\nu\right)} = \frac{\displaystyle \sum_{\left(m,n\right)\in\mathcal{L}} r^{\left(\nu-1\right)} \left[m,n\right] \varphi^\ast_{\left(k,l \right)}\left[m,n\right] w\left[m,n\right]}{\displaystyle \sum_{\left(m,n\right)\in\mathcal{L}} \varphi_{\left(k,l \right)}^\ast\left[m,n\right]w\left[m,n\right]\varphi_{\left(k,l \right)}\left[m,n\right]},
\end{equation}
as shown in \cite{Seiler2015}. In this context, the weighting function
\begin{equation}
\label{eq:weighting_function}
w\left[m,n\right]\hspace{-1mm} = \hspace{-1mm}\left\{ \hspace{-1mm}\begin{array}{ll} \hat{\rho}^{\sqrt{\left(m-\frac{M-1}{2}\right)^2+\left(n-\frac{N-1}{2}\right)^2}} & \mbox{for } \left(m,n\right)\in \mathcal{A} \\ \delta \hat{\rho}^{\sqrt{\left(m-\frac{M-1}{2}\right)^2+\left(n-\frac{N-1}{2}\right)^2}} & \mbox{for } \left(m,n\right)\in \mathcal{R} \\ 0 & \mbox{for } \left(m,n\right)\in \mathcal{B}\end{array}\right. ,
\end{equation}
is used for assigning an exponentially decreasing weight and therewith less influence on the model generation with increasing distance to the center block. The parameter $\hat{\rho}$ controls the decay. Since previously reconstructed samples are not as reliable as originally known ones, their influence is further attenuated by a factor $\delta$.

Based on the projection coefficients, the basis function is selected which reduces the weighted residual between the model and the available samples the most, subject to a frequency prior $w_f\left[k,l\right]$. As proposed in \cite{Seiler2015}, this prior assures that low frequencies are favored instead of high frequency ones and has the form
\begin{equation}
\label{eq:OTF-prior}
w_f\left[k,l\right] = \left( 1 - \sqrt{2} \sqrt{\frac{\tilde{k}^2}{M^2} + \frac{\tilde{l}^2}{N^2}} \right)^2
\end{equation} 
with $\tilde{k} = \frac{M}{2} - \left|k - \frac{M}{2}\right|$ and $\tilde{l} = \frac{N}{2} - \left|l-\frac{N}{2}\right|$. As shown in \cite{Seiler2015}, the prior can be regarded as an approximation of the Optical Transfer Function (OTF) of a diffraction limited optical system, which also attenuates high spatial frequencies more than low frequency ones.

Using the prior, the basis function to be selected results to 
\[
\left(u,v\right) = \argmax_{\left(k,l \right)} \Bigg( \left|p_{\left(k,l \right)}^{\left(\nu\right) }\right|^2 w_f\left[k,l\right] \cdot\hspace{2cm}
\]\vspace{-0.2cm}
\begin{equation}
\label{eq:bf_selection}
\hspace{1.5cm}\sum_{\left(m,n\right)\in\mathcal{L}}  \varphi_{\left(k,l \right)}^\ast\left[m,n\right]w\left[m,n\right]\varphi_{\left(k,l \right)}\left[m,n\right]\Bigg).
\end{equation}
After the basis function to be added in the current iteration $\nu$ has been selected, its weight is estimated by
\begin{equation}
\hat{c}_{\left(u,v \right)}^{\left(\nu\right)} = \gamma p_{\left(u,v \right)}^{\left(\nu\right)}.  
\end{equation}
The factor $\gamma$ is an orthogonality deficiency compensation factor \cite{Seiler2008} and is applied for obtaining a stable estimation. After the basis function selection and the weight estimation, the update of the model and the residual follows	
\begin{equation}
g^{\left(\nu\right)}\left[m,n\right] = g^{\left(\nu-1\right)}\left[m,n\right] + \hat{c}_{\left(u,v \right)}^{\left(\nu\right)} \varphi_{\left(u,v \right)}\left[m,n\right]
\end{equation}
\begin{equation}
r^{\left(\nu\right)}\left[m,n\right] = r^{\left(\nu-1\right)}\left[m,n\right] - \hat{c}_{\left(u,v \right)}^{\left(\nu\right)} \varphi_{\left(u,v \right)} \left[m,n\right].
\end{equation}
These steps are repeated until a predefined number of iterations is reached. Finally, the samples of the model corresponding to the block in the center are regarded as the reconstruction of the unknown samples. After one block has been finished, FSR proceeds to the next block, reusing the samples of the just reconstructed block for supporting the modeling of the next one.

\section{Adaptive Frequency Prior for Frequency Selective Reconstruction}
\label{sec:adaptive_prior}

Regarding the original FSR algorithm, it can be observed, that the frequency prior which controls the selection of the basis functions with respect to their spatial frequency is fixed and independent of the signal in reconstruction area $\mathcal{L}$. Accordingly, the same prior is used for the cases where $\mathcal{L}$ contains few known samples as well as for cases where almost all samples within $\mathcal{L}$ are known. However, in the case that many samples in $\mathcal{L}$ are known, there is still sufficient information available for reconstructing more high-frequent content whereas in the case that $\mathcal{L}$ only contains very few known samples, an even more low-frequency solution should be generated for avoiding artifacts. Thus, an adaptive frequency prior is introduced in the following which takes the local density of the available samples into account and adapts the modeling process to the given data.

The basic idea of the OTF-inspired prior from (\ref{eq:OTF-prior}) is to favor low-frequency basis functions over high-frequency ones during the selection process of FSR \cite{Seiler2015}. However, its influence on the selection process might be too strong or too weak, depending on the local density of the available data. 

For making the prior adaptive to the available samples, the measure
\begin{equation}
	\Omega = \frac{\displaystyle \sum_{\left(m,n\right)\in\mathcal{A}\cup\mathcal{R}} w\left[m,n\right]}{\displaystyle \sum_{\left(m,n\right)\in\mathcal{L}} \hat{\rho}^{\sqrt{\left(m-\frac{M-1}{2}\right)^2+\left(n-\frac{N-1}{2}\right)^2}}}
\end{equation}
of the effective data is introduced. The measure reuses the weighting function $w\left[m,n\right]$ from (\ref{eq:weighting_function}) in the same way as for the weighted projection: samples more far away from the block to be reconstructed should have a lower influence than the ones close to the block. Furthermore, samples that belong to neighboring already reconstructed blocks should have less influence than originally known ones. If $\hat{\rho}$ was one and all samples have same weight, the measure $\Omega$ of the effective data would just be the number of available samples with respect to the overall number of samples in $\mathcal{L}$.

\begin{figure}
	\centering
	\input{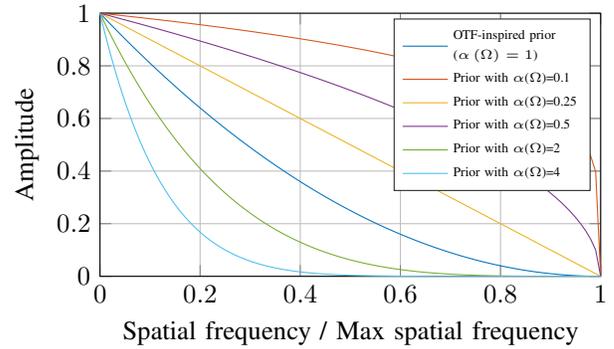}
	\caption{1D plot of frequency prior with OTF-inspired prior and adaptive frequency prior $w_f^{\left(\Omega\right)}\left[k,l\right]$ with different values of $\alpha\left(\Omega\right)$.\vspace{-0.1cm}}
	\label{fig:frequency_prior}
\end{figure}

Based on the measure $\Omega$, the frequency prior can be adapted. The objective is, if many samples are available, the suppression of high-frequency basis functions should be small, whereas in the opposite case, a strong suppression should be possible. This could be achieved by modifying the frequency prior to
\begin{equation}
\label{eq:apative_prior}
w_f^{\left(\Omega\right)}\left[k,l\right] = \left( 1 - \sqrt{2} \sqrt{\frac{\tilde{k}^2}{M^2} + \frac{\tilde{l}^2}{N^2}} \right)^{2 \alpha\left(\Omega\right)}
\end{equation}
for the basis function selection (\ref{eq:bf_selection}). Since the original OTF-inspired prior already achieved good results, the proposed adaptive prior of course is related to it, however, by introducing $\alpha\left(\Omega\right)$ in the exponent, its shape can be changed and its behavior can be made adaptive to the measure $\Omega$ of the effective data. The term $\alpha\left(\Omega\right)$ depicts a mapping function as the measure $\Omega$ of the effective data cannot be directly inserted into the exponent. In Fig.\ \ref{fig:frequency_prior}, a one-dimensional plot of the adaptive frequency prior is shown in comparison to the OTF-inspired prior. It can be observed that the OTF-inspired prior effectively favors low-frequency basis functions over high-frequency ones. However, it can also be discovered that for different values of $\alpha\left(\Omega\right)$, the behavior of the adaptive prior $w_f^{\left(\Omega\right)}\left[k,l\right]$ changes, as desired. That is to say, for small values of $\alpha\left(\Omega\right)$, the influence of the prior gets smaller, allowing also the selection of high-frequency basis functions whereas for large values of $\alpha\left(\Omega\right)$, the reconstruction of a low-frequency solution is favored.

Using this, the properties of the mapping function $\alpha\left(\Omega\right)$ can be formulated. For $\Omega$ close to $1$, that is to say when $\mathcal{L}$ contains many known samples, a reconstruction of high frequency content still is possible and the suppression of the high frequency basis functions should be small. This can be achieved if $\alpha\left(\Omega\right)$ goes towards $0$. In contrast to this, for small $\Omega$ which means that there is only little information available in $\mathcal{L}$, no reasonable high-frequency content can be reconstructed. Hence, the reconstruction of a low-frequency solution should be favored which can be achieved by $\alpha\left(\Omega\right)$ going towards $\infty$. Since the OTF-inspired prior achieved very good performances for the cases where between $10\%$ and $50\%$ of the samples are available as shown in \cite{Seiler2015}, the region where $\alpha\left(\Omega\right)$ becomes one should be in the same area, as well.

In order to fulfill these constraints, we propose to define the mapping function $\alpha\left(\Omega\right)$ of the adaptive frequency prior $w_f^{\left(\Omega\right)}\left[k,l\right]$ as
\begin{equation}
 \alpha\left(\Omega\right) = - \frac{\log\left(\Omega\right)}{\tau}
\end{equation}
with the parameter $\tau$ controlling for which $\Omega$ the adaptive prior becomes equal to the OTF-inspired prior. Defining the mapping function in this way, all the properties of the frequency prior formulated above can be achieved. At the same time, the mapping function is simple to calculate and only relies on one parameter to be determined. The next section now shows, how the reconstruction performance of FSR can be improved if the OTF-inspired prior from (\ref{eq:OTF-prior}) is replaced by the adaptive prior from (\ref{eq:apative_prior}).

\section{Simulations and Results}
\label{sec:results}

For showing the efficacy of the adaptive frequency prior, the reconstruction has been tested for a non-regular subsampling of images. That is to say, from an image, a non-regular subset of pixels has been selected and the objective always is to reconstruct the original image in the best possible way. For FSR, the parameters have been selected according to \cite{Seiler2015}, except for the frequency prior.

\begin{figure}
	\centering
%
%
\definecolor{mycolor1}{rgb}{0.00000,1.00000,1.00000}%
\definecolor{mycolor2}{rgb}{1.00000,0.00000,1.00000}%
\definecolor{mycolor3}{rgb}{1.00000,1.00000,0.00000}%
\begin{tikzpicture}

\begin{axis}[%
width=6.656cm,
height=3.75cm,
at={(0cm,0cm)},
scale only axis,
xmin=0,
xmax=100,
xlabel={Subsampling density $\left[\%\right]$},
xmajorgrids,
ymin=24,
ymax=47,
ylabel={PSNR [dB]},
ymajorgrids,
axis background/.style={fill=white},
legend style={at={(0.97,0.03)},anchor=south east,legend cell align=left,align=left,draw=white!15!black}
]
\addplot [color=blue,solid,mark=*,mark options={solid}]
  table[row sep=crcr]{%
5	24.6381946124901\\
10	26.084932489794\\
20	27.9042982176054\\
30	29.3711365856376\\
40	30.8118350924384\\
50	32.3625289516078\\
60	34.1173475701885\\
70	36.2062491156897\\
80	38.8836883049756\\
90	42.7752209296715\\
95	46.0762176709754\\
};
\addlegendentry{\tiny $\tau\text{=0.1}$};

\addplot [color=green,solid,mark=o,mark options={solid}]
  table[row sep=crcr]{%
5	24.5791793741634\\
10	26.3694945891347\\
20	28.8441829400093\\
30	30.7822055994891\\
40	32.4960928768382\\
50	34.1442803672552\\
60	35.8285826189558\\
70	37.6557632071724\\
80	39.9117241847337\\
90	43.2447837279151\\
95	46.2747574987153\\
};
\addlegendentry{\tiny $\tau\text{=0.5}$};

\addplot [color=red,solid,mark=x,mark options={solid}]
  table[row sep=crcr]{%
5	24.9533390261031\\
10	26.736463937767\\
20	29.1992644401574\\
30	31.1120564556663\\
40	32.7789750854151\\
50	34.3558664947474\\
60	35.9740637890529\\
70	37.7391655636818\\
80	39.9548294946873\\
90	43.2559708143924\\
95	46.2841569976829\\
};
\addlegendentry{\tiny $\tau\text{=1.0}$};

\addplot [color=mycolor1,solid,mark=+,mark options={solid}]
  table[row sep=crcr]{%
5	25.0774276680831\\
10	26.8777221588446\\
20	29.3063132307813\\
30	31.1817091760828\\
40	32.8092161182811\\
50	34.3616645318084\\
60	35.9672124580917\\
70	37.7306017046895\\
80	39.9491630414282\\
90	43.2567739071106\\
95	46.284573477412\\
};
\addlegendentry{\tiny $\tau\text{=1.5}$};

\addplot [color=mycolor2,solid,mark=asterisk,mark options={solid}]
  table[row sep=crcr]{%
5	25.1188173634759\\
10	26.9314810840131\\
20	29.331918273779\\
30	31.1804842151119\\
40	32.7905045128188\\
50	34.3373858256099\\
60	35.9431024788502\\
70	37.7149764757213\\
80	39.9412585606498\\
90	43.253833355578\\
95	46.2883413378337\\
};
\addlegendentry{\tiny $\tau\text{=2.0}$};


\addplot [color=black,solid,mark=diamond,mark options={solid}]
  table[row sep=crcr]{%
5	25.128569045318\\
10	26.9460935577244\\
20	29.3185577506472\\
30	31.136745194239\\
40	32.7308234721418\\
50	34.2768111275494\\
60	35.900200852933\\
70	37.6890778764463\\
80	39.9271143035085\\
90	43.2523501069699\\
95	46.2864862916117\\
};
\addlegendentry{\tiny $\tau\text{=3.0}$};


\end{axis}
\end{tikzpicture}%
\vspace{-2.5mm}
	\caption{Average reconstruction quality for Kodak dataset in dB PSNR with respect to the subsampling density for different values of parameter $\tau$.}
	\label{fig:threshold_determination}
\end{figure}
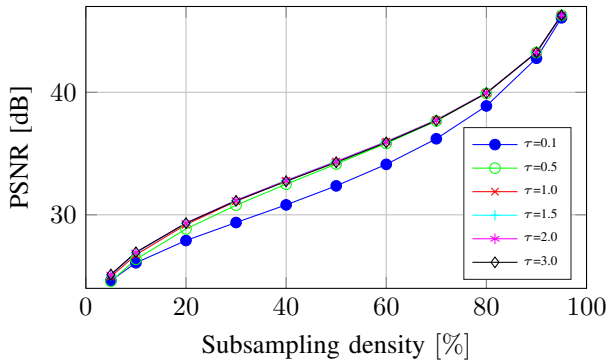

In order to determine a good threshold $\tau$ for the new prior $w_f^{\left(\Omega\right)}\left[k,l\right]$, tests have been carried out using the Kodak test database \cite{Kodak_Test_Data}. Fig.\ \ref{fig:threshold_determination} shows the reconstruction quality in PSNR averaged over the whole database, for different values of $\tau$, with respect to the subsampling density. The latter is the ratio of available samples compared to the number of samples on the regular grid. According to these tests, it can be observed, that except for very small values, the exact value of $\tau$ is not very critical and a threshold of $\tau=2$ is a useful choice.

In order to show that the proposed adaptive prior is superior to the fixed OTF-inspired prior, further tests have been carried out on the \mbox{TECNICK} image database \cite{Asuni2011}. This independent database has been selected in order to avoid a fitting of the parameter $\tau$ to the test data. In addition to the proposed adaptive frequency prior, denoted by FSR-AP, FSR has been tested with the original OTF-inspired prior, denoted by \mbox{FSR-OTF}. Furthermore, a reconstruction by several algorithms from literature has been performed for obtaining a comparison to the state-of-the-art. The considered algorithms are Linear Interpolation (LIN), Natural Neighbor Interpolation (NAT) \cite{Sibson1981}, Steering Kernel Regression (SKR) \cite{Takeda2007}, Sparse Wavelet Inpainting (SI) \cite{Starck2010}, and Constrained Split Augmented
Lagrangian Shrinkage Algorithm (CLS) \cite{Afonso2011}.

\begin{table}[t]
	\centering 
	\caption{Average processing time for the reconstruction of one image from the TECNICK dataset at a subsampling density of 10\%.}
	\label{tab:processing_time}
	
	\begin{tabular}{|c|c|c|c|c|c|c|}
		\hline FSR-AP & FSR-OTF & LIN & NAT & SKR & SI& CLS  \\ \hline 
		$448\ \mathrm{ s}$ & $476\ \mathrm{ s}$ & $5.4\ \mathrm{ s}$ & $6.4\ \mathrm{ s}$  & $193\ \mathrm{ s}$ & $2486\ \mathrm{ s}$ & $82\ \mathrm{ s}$ \\ \hline
	\end{tabular} 
\end{table}

\begin{figure}
	\centering
%
%
\definecolor{mycolor1}{rgb}{0.87059,0.49020,0.00000}%
\definecolor{mycolor2}{rgb}{0.00000,1.00000,1.00000}%
\definecolor{mycolor3}{rgb}{1.00000,0.00000,1.00000}%
\begin{tikzpicture}

\begin{axis}[%
width=6.656cm,
height=3.75cm,
at={(0cm,0cm)},
scale only axis,
xmin=0,
xmax=100,
xlabel={Subsampling density $\left[\%\right]$},
xmajorgrids,
ymin=0,
ymax=3.5,
ylabel={PSNR gain [dB]},
ymajorgrids,
axis background/.style={fill=white},
legend style={at={(0.97,0.03)},anchor=south east,legend cell align=left,align=left,draw=white!15!black}
]

\addplot [color=red,solid,mark=asterisk,mark options={solid}]
  table[row sep=crcr]{%
5	0.819737459469109\\
10	1.43318912436798\\
20	2.06290879374091\\
30	2.38565133519767\\
40	2.57743311784226\\
50	2.68589610317985\\
60	2.76324480422617\\
70	2.84360143263834\\
80	2.93817764469393\\
90	3.0152452740718\\
95	2.93526611444236\\
};
\addlegendentry{\tiny FSR-AP};

\addplot [color=green,dashed,mark=o,mark options={solid}]
table[row sep=crcr]{%
	5	0.810907791144632\\
	10	1.44524179356375\\
	20	2.02002561648104\\
	30	2.27477657550815\\
	40	2.38721331286405\\
	50	2.4063003860699\\
	60	2.38738141335569\\
	70	2.37042583245156\\
	80	2.37555106445922\\
	90	2.38733492935453\\
	95	2.31467989169971\\
};
\addlegendentry{\tiny FSR-OTF \cite{Seiler2015}};

\addplot [color=mycolor1,solid,mark=*,mark options={solid}]
  table[row sep=crcr]{%
5	0.254517062535321\\
10	0.236670351358457\\
20	0.239402942642942\\
30	0.268749447317525\\
40	0.319560788111637\\
50	0.377265078016896\\
60	0.454362462723726\\
70	0.575717315406523\\
80	0.76732047616103\\
90	1.04861356413158\\
95	1.16485740783537\\
};
\addlegendentry{\tiny NAT \cite{Sibson1981}};

\addplot [color=blue,solid,mark=x,mark options={solid}]
  table[row sep=crcr]{%
5	-9.52117793239018\\
10	-0.144935960169489\\
20	1.04437211075633\\
30	0.677956829101777\\
40	0.204532072179591\\
50	-0.237778234347562\\
60	-0.618319514149147\\
70	-0.909270646227889\\
80	-1.10339530333794\\
90	-1.18509676091985\\
95	-1.18709395339917\\
};
\addlegendentry{\tiny SKR \cite{Takeda2007}};

\addplot [color=mycolor2,solid,mark=square,mark options={solid}]
  table[row sep=crcr]{%
5	-1.90763089997452\\
10	-1.08568852000429\\
20	0.107571902173405\\
30	0.759578891989825\\
40	1.24545036392453\\
50	1.62292094298333\\
60	1.95220394464416\\
70	2.27091460967855\\
80	2.58394779648869\\
90	2.8685420840098\\
95	2.84712230195485\\
};
\addlegendentry{\tiny SI \cite{Starck2010}};

\addplot [color=mycolor3,solid,mark=diamond,mark options={solid}]
  table[row sep=crcr]{%
5	-1.38896968394525\\
10	-1.31773361687746\\
20	-0.914075985004963\\
30	-0.491548661621142\\
40	-0.0604710313016472\\
50	0.384190071159606\\
60	0.844959397603243\\
70	1.35489557404897\\
80	1.92460536663931\\
90	2.55113019327035\\
95	2.761795041682\\
};
\addlegendentry{\tiny CLS \cite{Afonso2011}};

\end{axis}
\end{tikzpicture}%
\vspace{-2.5mm}
	\caption{Average gain in reconstruction quality for TECNICK dataset in dB PSNR compared to a reconstruction by Linear Interpolation for different subsampling densities.}
	\label{fig:reconstruction_quality_gain}
\end{figure}
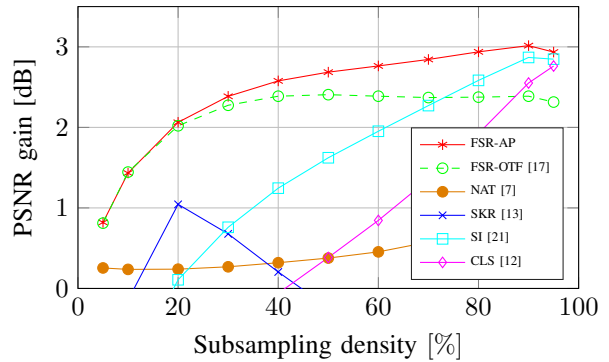

\begin{table*}[t]
	\centering 
	\caption{Reconstruction quality for TECNICK dataset in dB PSNR for different subsampling densities.}
	\label{tab:reconstruction_quality}
	
	\begin{tabular}{|l|c|c|c|c|c|c|c|c|c|c|c|}
		\hline Subsampling density& $5\%$ & $10\%$ & $20\%$ & $30\%$ & $40\%$ & $50\%$ & $60\%$ & $70\%$ & $80\%$ & $90\%$ & $95\%$ \\ \hline
		\hline FSR-AP & $\bf 27.03$ & $29.69$ & $\bf 32.83$ & $\bf 34.94$ & $\bf 36.66$ & $\bf 38.23$ & $\bf 39.79$ & $\bf 41.50$ & $\bf 43.59$ & $\bf 46.75$ & $\bf 49.55$ \\
		\hline FSR-OTF \cite{Seiler2015} & $27.02$ & $\bf 29.70$ & $32.78$ & $34.83$ & $36.47$ & $37.95$ & $39.42$ & $41.03$ & $43.03$ & $46.12$ & $48.93$ \\
		\hline LIN & $26.21$ & $28.26$ & $30.76$ & $32.55$ & $34.09$ & $35.54$ & $37.03$ & $38.66$ & $40.65$ & $43.73$ & $46.61$ \\
		\hline NAT \cite{Sibson1981} & $26.46$ & $28.49$ & $31.00$ & $32.82$ & $34.41$ & $35.92$ & $37.48$ & $39.23$ & $41.42$ & $44.78$ & $47.78$ \\
		\hline SKR \cite{Takeda2007} & $16.69$ & $28.11$ & $31.81$ & $33.23$ & $34.29$ & $35.30$ & $36.41$ & $37.75$ & $39.55$ & $42.55$ & $45.42$ \\
		\hline SI \cite{Starck2010} & $24.30$ & $27.17$ & $30.87$ & $33.31$ & $35.33$ & $37.16$ & $38.98$ & $40.93$ & $43.23$ & $46.60$ & $49.46$ \\
		\hline CLS \cite{Afonso2011} & $24.82$ & $26.94$ & $29.85$ & $32.06$ & $34.03$ & $35.92$ & $37.87$ & $40.01$ & $42.58$ & $46.28$ & $49.37$ \\ \hline 
	\end{tabular} 
\end{table*}

\begin{figure*}[t]
\hspace{-1cm}
	\psfrag{Original}[c][c][1][90]{Original}
	\psfrag{Subsampled}[c][c][1][90]{Subsampled}
	\psfrag{SI}[c][c][1][90]{SI \cite{Starck2010}}
	\psfrag{SKR}[c][c][1][90]{SKR \cite{Takeda2007}}
	\psfrag{CLS}[c][c][1][90]{CLS \cite{Afonso2011}}
	\psfrag{FSR-OTF}[c][c][1][90]{FSR-OTF \cite{Seiler2015}}
	\psfrag{FSR-AP}[c][c][1][90]{FSR-AP}
	\includegraphics[width=0.99\textwidth]{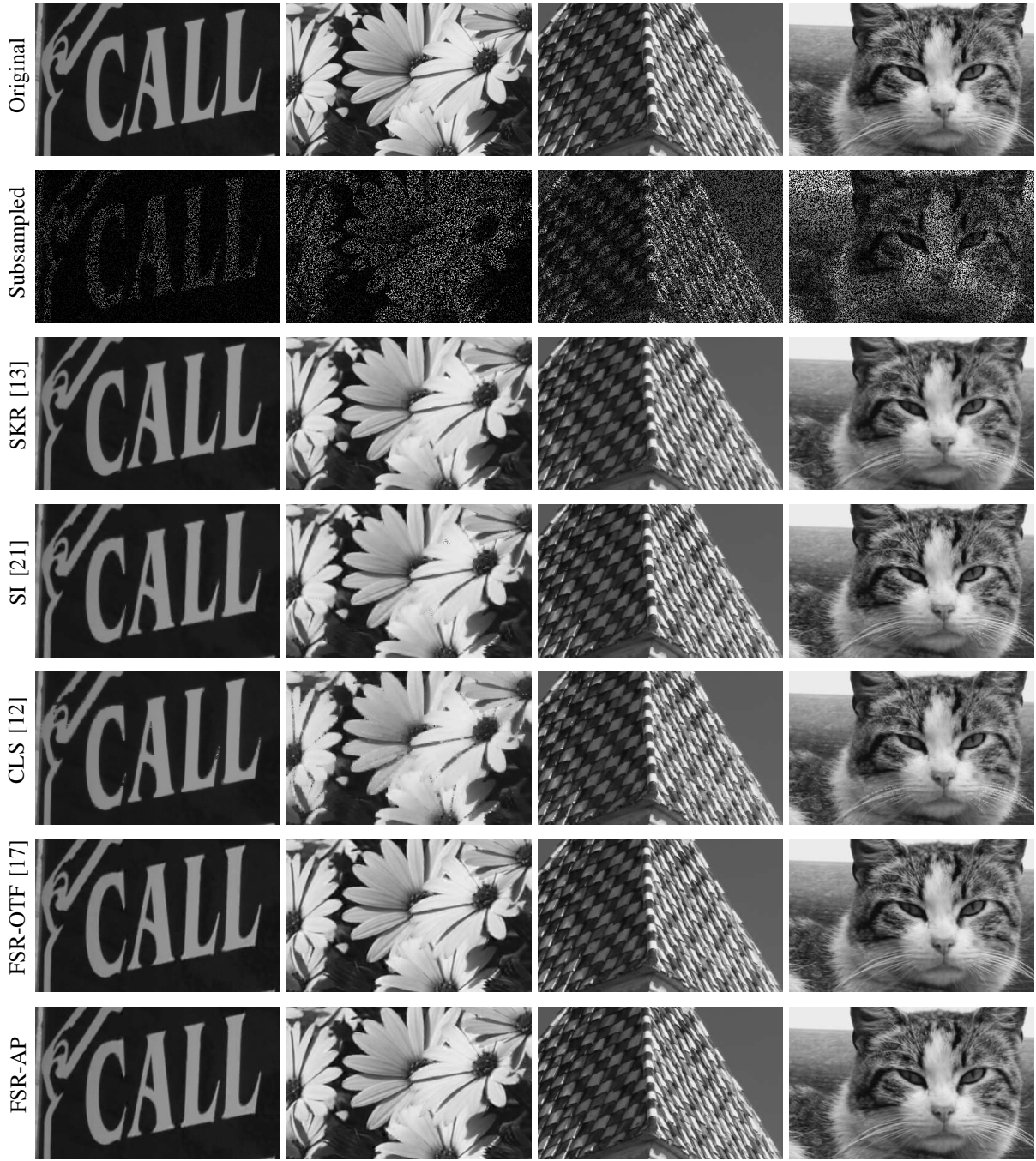}
	\caption{Visual reconstruction results for details of four images of the TECNICK dataset with a subsampling density of $20\%$ (first two columns) and $50\%$ (last two columns). \emph{(Please pay attention, additional aliasing may be caused by printing or scaling. Best to be viewed enlarged on a monitor.)}}
	\label{fig:visual_results}
\end{figure*}

In Table \ref{tab:reconstruction_quality}, the average reconstruction quality is listed for the considered algorithms and for different subsampling densities. Furthermore, Fig.\ \ref{fig:reconstruction_quality_gain}, shows the average gain with respect to a reconstruction by Linear Interpolation. Examining the results, it can be observed that the OTF-inspired prior \cite{Seiler2015} only performs superior for low subsampling densities, that is to say, when only few samples are available. For high subsampling densities, some of the other algorithms perform better. However, the proposed adaptive prior assures that FSR achieves a very high reconstruction quality over the whole range of subsampling densities. In doing so, FSR-OTF with the OTF-inspired prior is outperformed by $0.6$ dB for high subsampling densities. In addition to this, the adaptive prior leads to the highest reconstruction quality over the whole range of subsampling densities (only beaten by negligible $0.01$ dB by the OTF-inspired prior for a subsampling density of $10\%$). Compared to the state-of-the-art algorithms from literature, it can be observed, that FSR with the proposed adaptive frequency prior is able to outperform the other algorithms independent of the number of available samples and gains of several dB in PSNR are possible. 

Of course, introducing the adaptive frequency prior slightly increases the complexity of the reconstruction. In order to asses the impact, Table \ref{tab:processing_time} lists the average processing time for the reconstruction of one image from the TECNICK database at a subsampling density of 10\%. It can be observed that the calculation of the adaptive prior increases the complexity of FSR only by $6\%$. Given the improved quality, this small increase of the complexity is a reasonable trade-off.

In order to show that the gains are also visually noticeable, Fig.\ \ref{fig:visual_results} shows details of four images from the TECNICK image database. In this figure, the output of the reconstruction by FSR with the proposed adaptive prior and some of the state-of-the-art algorithms is shown. It can be observed that the proposed algorithm yields the highest visual quality with only very little artifacts, well reflecting the PSNR results.

\section{Conclusion}
\label{sec:conclusion}

In this paper an adaptive frequency prior was introduced for improving Frequency Selective Reconstruction of image signals from a non-regular subsampling. For this, the local density of the available image samples is taken into account and depending on this, the prior is adapted to favor the reconstruction of high- or low-frequency content. While the fixed frequency prior of the original Frequency Selective Reconstruction only performs superior for the case that few samples are available, the proposed adaptive frequency prior allows for a very high reconstruction quality, independent of the number of available samples. Compared to the fixed prior, gains of up to $0.6$ dB can be achieved. Furthermore, other state-of-the-art algorithms can be outperformed by several dB PSNR which can also be visually observed.
\clearpage
\balance


\end{document}